**RaspberryPi to measure EEG, ECG, EMG and EOG with Shield PiEEG**

In this paper presented hardware and software for shield PiEEG for reading signals through the families of single-board computers - RaspberryPi, OrangePi, BananaPi, etc. For the most part, the paper provides technical information on how to implement this device. This device is designed to be familiar with neuroscience and is one of the easiest ways to get started with EEG measurements.

Ildar Rakhmatulin, PhD, PiEEG, ildarr2016@gmail.com
Source https://github.com/Ildaron/EEGwithRaspberryPI
Presentation https://youtu.be/uK8QF2liO5U

## 1. Introduction

The brain-computer interface is a device for reading brain signals in order to identify any correlations that can be used for practical purposes. In 2021 we developed the brain-computer interface - ironbci [1,2,3] but the chip shortage significantly increased the cost of the device, after which we switched to PiEEG shield, which made it possible to reduce the cost of the device and simplify the installation process. The PiEEG device was presented in general terms at the conference [4] and in the publication [10]. In this paper we will focus more on the technical details of the implementation of this device.

## 2. Safety recommendation

The developed device was tested only for Raspberry Pi. During the test, it is forbidden connect the device to the electrical source, this need for safety and to avoid network interference. Cannot be used this device when powered via an electrical network and used it only when working with 5V batteries (capacity– not mote 2000 mAh). General view of the complete assembly of the device in Fig.1.

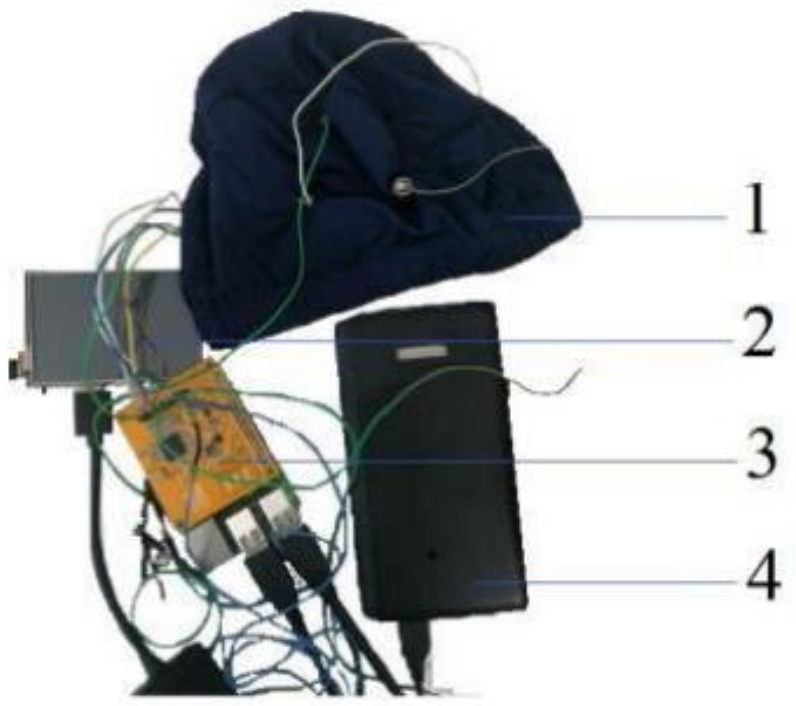

Fig.1 General View of assembled device: 1 – hat EEG, 2 – monitor, 3 – RaspberryPI and shield EEG, 4 – power bank

All parts should be connected only with battery – monitor, keyboard, mouse, RaspberryPI and PiEEG device. The device is connected only directly through GPIO40 to the Raspberry Pi. There must not be any connections for any elements with the electrical network.
Also observe the following protective measures:

- If the device fails (heating of components, smoke etc. ), switch off immediately;
- When you have finished using any electrical appliance, be sure to turn it off and unplug it;
- Never try to fix a malfunction in an electrical appliance yourself;
- Do not touch the components on the board when the device is turned on.

## 3. Connecting the device

The device is designed to allow single-board computers to read the EEG signal via the SPI protocol from ADS1299. The next single computers can be connected (not tested yet) - OrangePi, BananaPi, and the device can be connected via wires to the input of the microcontroller if necessary, Fig.2.

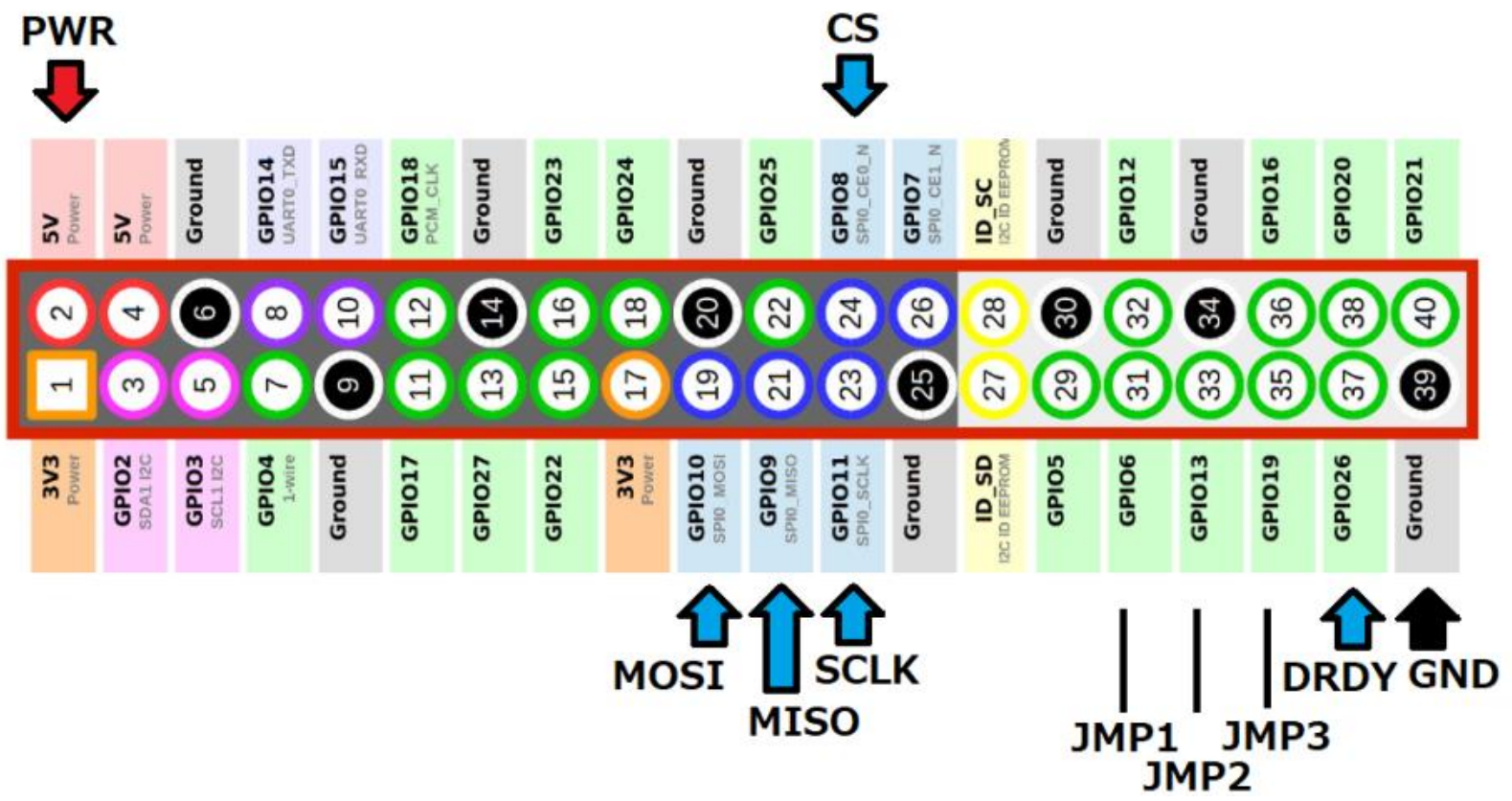

Fig.2. Device connection scheme, GPIO40

**4. Software**

Data transfer between Raspberry Pi and ADS1299 using the SPI protocol is written in C at a rate of 250 samples per second for each channel. Data processing realized in Python 3. Software files details from https://github.com/Ildaron/EEGwithRaspberryPI/tree/master/GUI :

libperiphery.a – static library compiled from https://github.com/vsergeev/c-periphery ,
real_time.py – script for editing processing signal and data visualization,
real_time_massive.c – header call,
real_time_massive.h – register setting for data transfer protocol,
super_real_time_massive.so – compiled dynamic library.

To start the reading EEG signals is possible to write software or download from the git https://github.com/Ildaron/EEGwithRaspberryPI , then go to the folder - GUI and run the file - real_time.py. An example of reading is shown in the figure in Fig.3.

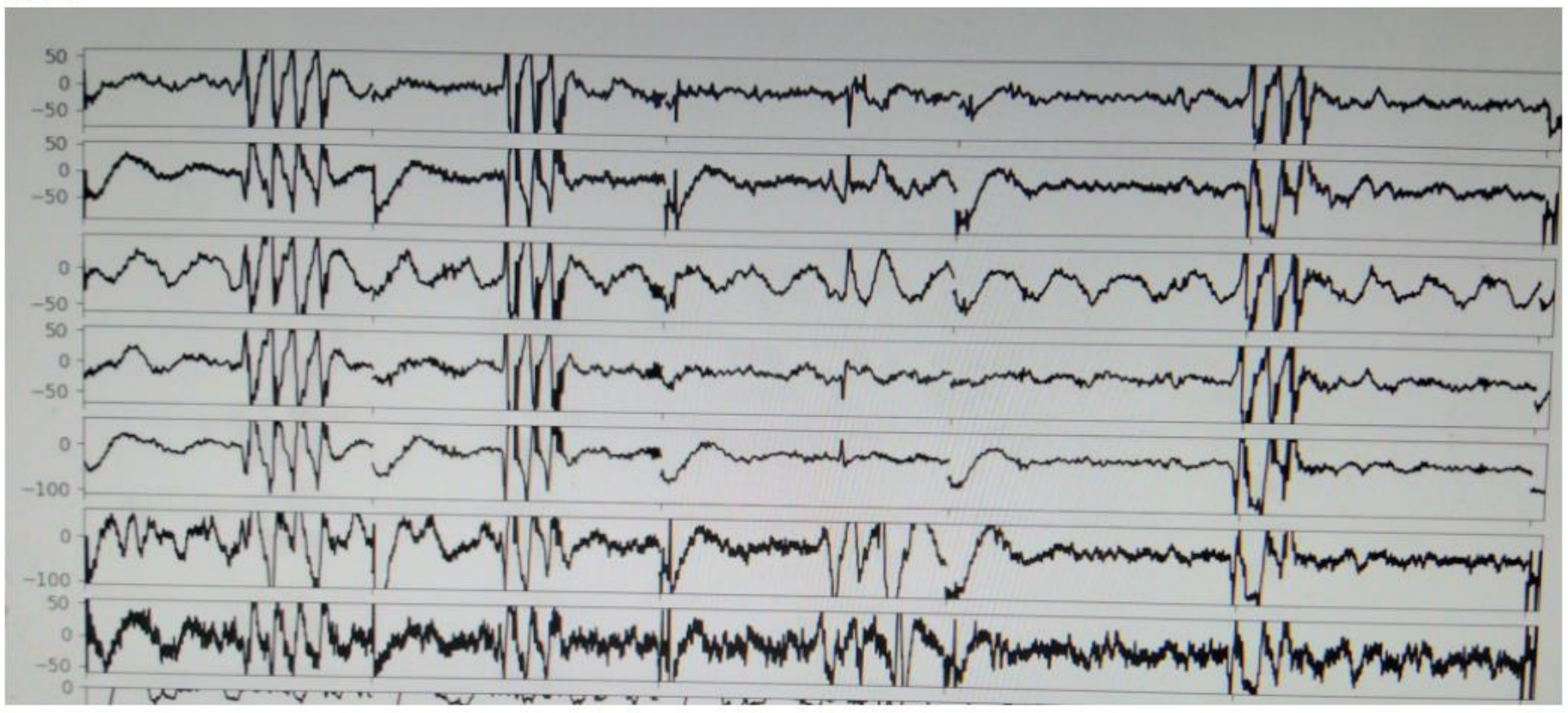

Fig.3. EEG signals after band pass filter 1-30 Hz

Through the python script, is possible to configure the parameters for band pass filters. Signal processing is implemented using the standard scipy library. Through the C file, once per second we receive a data packet of 2000 data, 250 samples per second for each channel.

Currently we used the next setting for registers for ADS1299:

```
write_reg (0x14,0x80);//led
write_reg (0x05,0x00);//ch1
write_reg (0x06,0x01);//ch2
write_reg (0x07,0x01);//ch3
write_reg (0x08,0x01);//ch4
 write_reg (0x0D,0xFF);//bias N
 write_reg (0x0E,0x00);//bias P
write_reg (0x09,0x01);//ch5
write_reg (0x0A,0x01);//ch6
write_reg (0x0B,0x01);//ch7
write_reg (0x0C,0x01);//ch8
```

```
write_reg (0x15,0x20);// mics
write_reg (0x01,0x96);// reg1
write_reg (0x02,0xD4);// reg2
write_reg (0x03,0xE0);// reg3
send_command (0x10);//sdatc
send_command (0x08);//start
```

To change the ADS1299 registers, needs to work with the header file - real_time_massive.h, after any changes, need to recompile the static library, following the next steps for GPIO library for Linux from https://github.com/vsergeev/c-periphery:

```
$ mkdir build
$ cd build
$ cmake ..
$ make
```

That allows receiving the Static Library - libperiphery.a and finally to create a dynamic library for .so - gcc -shared -home/pi/Desktop/new_spi/c-periphery-master/src **real_time_massive.c** /home/pi/Desktop/new_spi/c-periphery-master/build/**libperiphery.a** -o **super_real_time_massive.so**

This library can be imported in python script with signal processing (example - real_time.py) via standard library – ctypes.

In case of errors incompatibility in a python IDE, it is necessary to check the versions of installed libraries https://github.com/Ildaron/EEGwithRaspberryPI/blob/master/pip.txt

The PIEEG device has a diode connected to the ADS1299, which lights up when the PiEEG device is initialized. If the life bit is not burning, it is necessary to check the power (Fig.4) in the next points:

TP1: – 2.5V
TP2: 3.3V
TP3: 2.5V

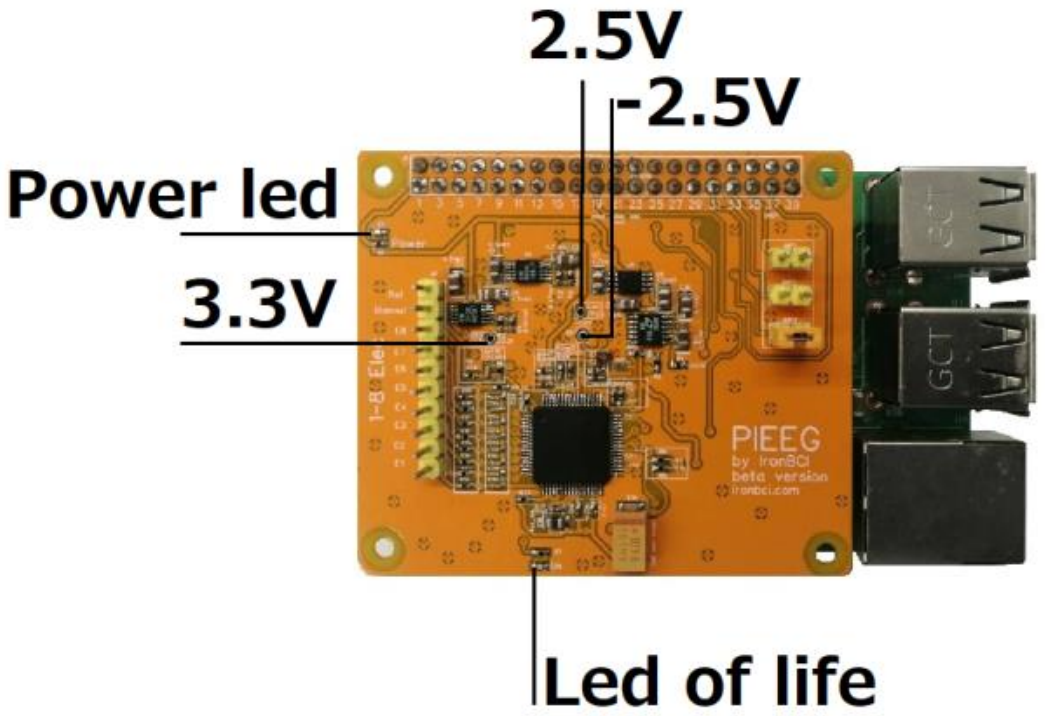

Fig.4. Test points in the PiEEG board

If there is noise in the measured EEG signals that goes beyond the limits declared in the device it is necessary to check the voltage in the next points:

VCAP1 = -1.2V
VCAP2 = 0.001V
VCAP3 = 4.2V
VCAP4=-0.125V

## 5. Conclusion and perspectives

This device is positioned as an easy way to start measuring EEG signals. We introduced hardware and software for reading and processing data in real-time in open source format. We demonstrated how to control a robot using a device [5], in the future we will work on the P300 and on other paradigms, then we plan to gather a team of like-minded people and collect the necessary amount of data to control robots through the motor imagery method. Fig.5 shows a block diagram for the possibility areas of use of the device.

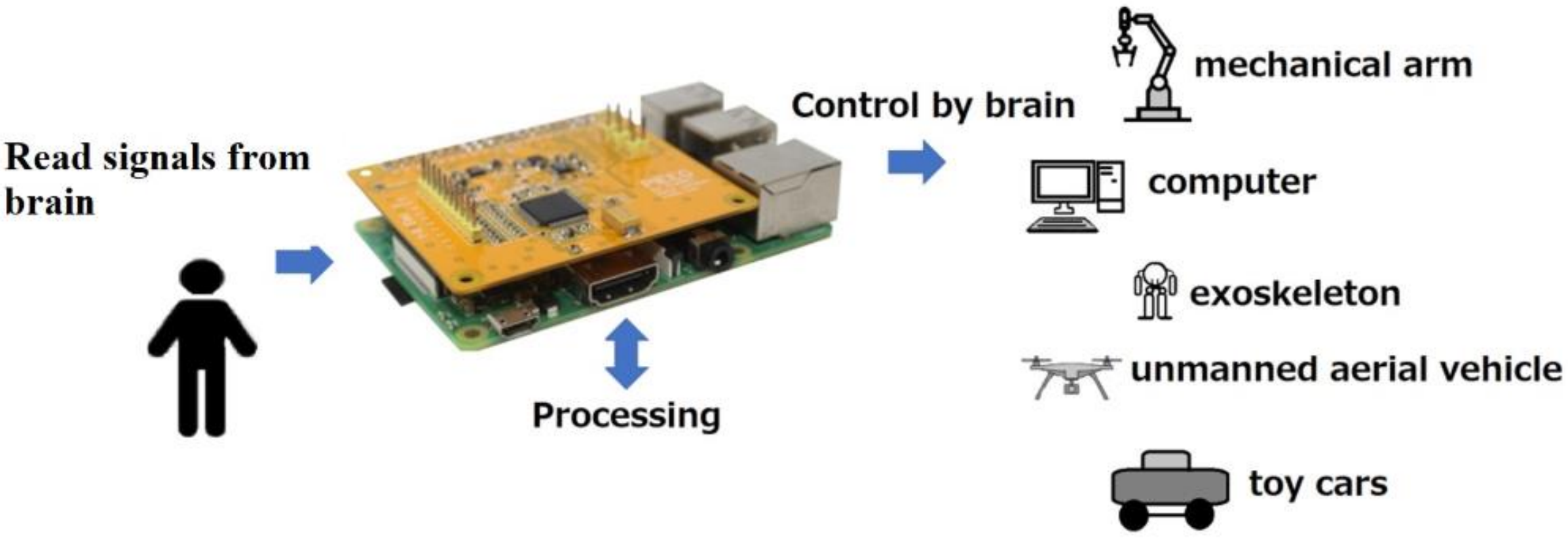

Fig.5. possibility areas of use of the PiEEG

**Additional Information**

Machine learning to identify alcoholism [9],
Overview of dry electrodes [6],
Signal processing [7,8] .